\documentclass[pra,twocolumn,10pt,aps]{revtex4-1}
\usepackage[dvips]{graphics,graphicx}
\usepackage{amsmath}

\usepackage{amssymb}
\usepackage[usenames, dvipsnames]{color}
\usepackage{graphicx}
\usepackage{graphics}
\usepackage{hyperref}

\usepackage{amsfonts}
\usepackage{eufrak}
\usepackage{mathrsfs}
\usepackage{bm}

\usepackage{soul}
\usepackage{soulutf8}

\newcommand{\scs}{\scriptscriptstyle}
\newcommand{\smallm}{\scs (\!-\!)}
\newcommand{\smallp}{\scs (\!+\!)}

\newcommand{\smallpm}{\scs (\!\pm\!)}

\begin{document}

\title{Fabry--Perot Bound States in the Continuum in an Anisotropic Photonic Crystal}
\author{Stepan~V.~Nabol, Pavel~S.~Pankin$^{*}$, Dmitrii~N.~Maksimov and Ivan~V.~Timofeev}
\affiliation{Kirensky Institute of Physics, Krasnoyarsk Scientific Center, Siberian Branch, Russian Academy of Sciences, Krasnoyarsk, 660036 Russia}
\affiliation{Siberian Federal University, Krasnoyarsk, 660041 Russia}
\affiliation{$^{*}$pavel-s-pankin@iph.krasn.ru}

\date{\today}
\begin{abstract}

An anisotropic photonic crystal containing two anisotropic defect layers is considered. It is demonstrated that the system under can support a Fabry--Perot bound state in the continuum (FP-BIC). A fully analytic solution of the scattering problem as well as a condition for FP-BIC have been derived in the framework of the temporal coupled-mode theory.

\end{abstract}
\maketitle

\section{Introduction}

The bound state in the continuum (BIC) is a nonradiative eigenstate of an open system the eigenvalue of which lies in the continuum of propagating waves \cite{HsuChiaWei2016, koshelev2022bound, azzam2021photonic, joseph2021bound}. The BICs were first found when solving the problem of the eigenenergy of a particle in a spherical quantum well \cite{Neumann}. Von Neumann and Wigner found special oscillating potentials tending asymptotically to zero far from the quantum well, the destructive interference on which allows a particle to stay localized even at the energies above the potential well. The BIC is a general wave phenomenon, which is always caused by the destructive interference of waves leaking from a system. For convenience, the BICs are classified, according to types of their formation, into symmetry-protected, Friedrich--Wintgen, Fabry--Perot, and accidental \cite{sadreev2021interference}.

Theoretically, the BICs have an infinite Q-factor, since they do not radiate into the environment. To excite and detect the resonance, the BIC should be coupled with the propagating waves. Then, the BIC turns into a quasi-BIC with a finite Q-factor. Varying the parameters of a system near the BIC, one can control the coupling between the resonance and the continuum, i.e., the resonant Q-factor. The quasi-BICs with controllable Q-factor have been proposed for various photonics applications, e.g., lasers \cite{Kodigala17, hwang2021ultralow, yang2021low}, light filters \cite{hu2022high, abujetas2021high, doskolovich2019integrated}, sensors \cite{Romano2019_BICsensor, maksimov2022enhanced, huo2022highly}, waveguides \cite{vega2021qubit, ye2022second, Bezus2018_BIC, ovcharenko2020bound}, and amplification of nonlinear effects   \cite{bernhardt2020quasi, liu2021giant, carletti2019high}.

The BICs can be implemented in three-, two-, and one-dimensional structures extended at least in one spatial dimension \cite{HsuChiaWei2016}. The BICs in a one-dimensional photonic structure were first implemented in \cite{GomisBresco_Torner2017_BIC1D}, where the authors created a trilayer waveguide made of anisotropic Materials, which supported a BIC with a theoretically infinite path length. In one-dimensional structures based on photonic crystals (PhCs) with anisotropic layers, BICs were studied theoretically \cite{Timofeev2018_BIC, Pankin2020Fano, pankin2022bound, ignatyeva2020bound} and experimentally  \cite{pankin2020one, wu2021quasi}.

In this work, we consider a one-dimensional anisotropic PhC consisting of alternating isotropic and anisotropic layers. Introducing one anisotropic defect layer in this PhC, one can engineer symmetry-protected BICs \cite{Timofeev2018_BIC, Pankin2020Fano} as well as Friedrich--Wintgen BICs \cite{pankin2022bound}. Here, we study the case of two anisotropic defect layers, which allow us to set up a Fabry--Perot BIC. Since each separate defect can exhibit a BIC-induced resonance, they are equivalent to two perfectly reflecting mirrors arranged in such a way that the waves are reflected in antiphase and compensate each other making it possible to obtain a Fabry--Perot BIC \cite{bulgakov2010bound, ndangali2010electromagnetic,huang2022topological,bulgakov2022desktop,sadreev2005trapping,sadreev2005trapping,bulgakov2010bound}.

\section{Model}

\begin{figure*}[t]
\center{\includegraphics{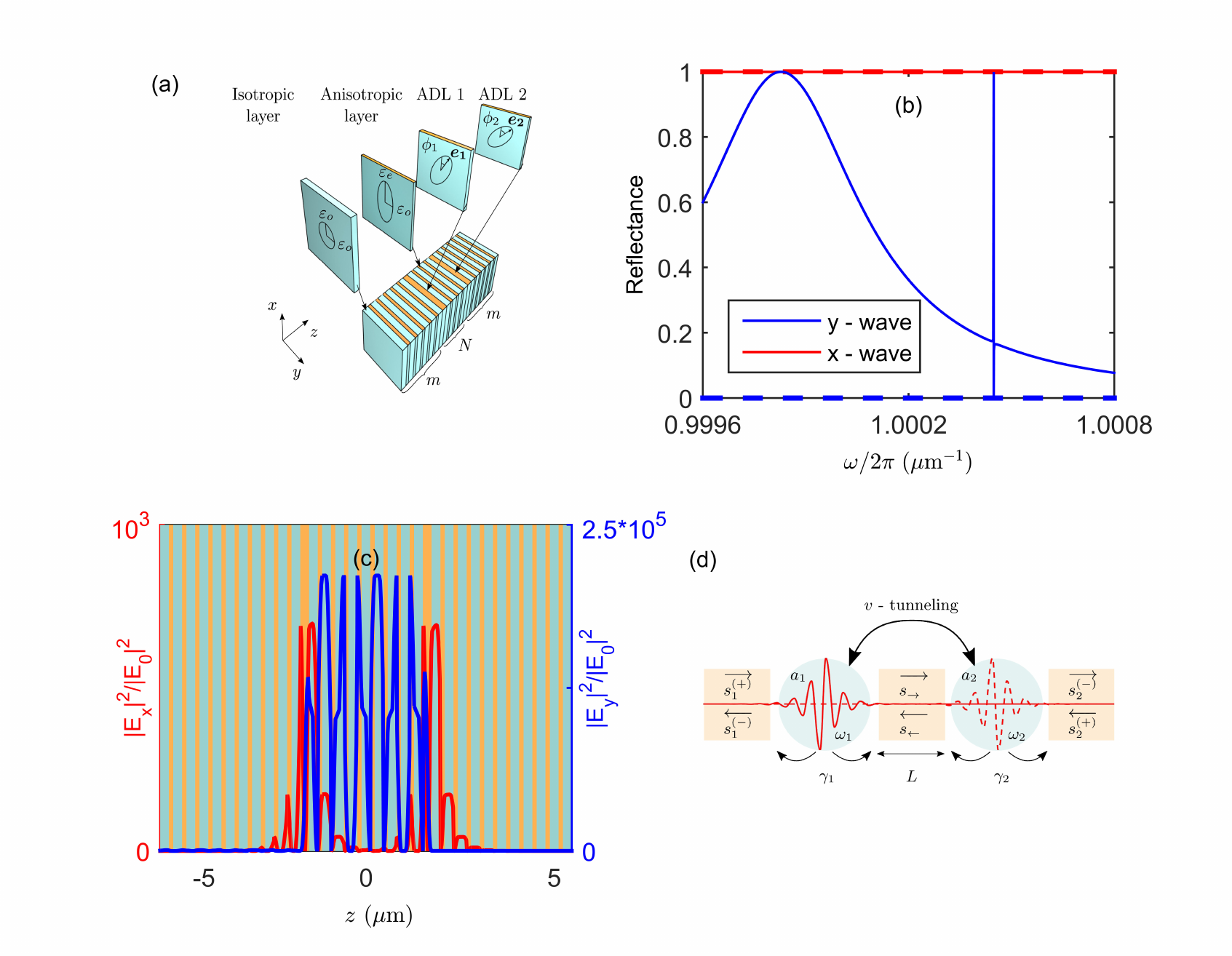}}
\caption{(a) Model of an anisotropic PhC with two defects. (b) Reflectance of the PhC for $y$-waves (blue) and $x$-waves (red) calculated by the Berreman method. The parameters are $n_o = 1$, $d_o = 0.25$ $\mu m$, $n_e = 2$, $d_e = 0.125$ $\mu m$, $N = 8$, $m = 20$, $\phi_1 = \phi_2  = 0$ (dashed line) and $\phi_1 = 2.4\pi/180$, $\phi_2  = 2.5\pi/180$ (solid line). (c) Distributions of the local field intensity $|\bm{E_{x,y}}|^2/|\bm{E}_0|^2$ at a frequency of $\omega/2\pi = 1.00045$ $\mu m^{-1}$ corresponding to the resonance in (b). (d) TCMT model of an anisotropic PhC with two defects.}
\label{fig1}
\end{figure*}

The model under scrutiny is a one-dimensional PhC consisting of alternating isotropic and anisotropic layers with two anisotropic defect layers (ADLs), ADL~1 and ADL~2, see Fig.~\ref{fig1}~(a). The refractive index of the isotropic layer is $n_o = \sqrt{\varepsilon_o}$ and the layer thickness is $d_o$. The anisotropic layer with thickness $d_e$ has the ordinary refractive index $n_o = \sqrt{\varepsilon_o}$ and the extraordinary refractive index $n_e = \sqrt{\varepsilon_e}$ for the waves polarized along the $y$-axis ($y$-wave) and $x$-axis ($x$-wave) direction, respectively. The layer thicknesses are quarter-wave and are determined by the equation
\begin{equation}
k_{o}d_o = k_{e}d_e = \frac{k_0 \lambda_{\scs \mathrm{PBG}}}{4} = \frac{\pi \omega}{2 \omega_{\scs \mathrm{PBG}}},
\label{QW}
\end{equation}
where $k_{o,e} = k_0 n_{o,e}$, $k_0 = \omega/c$ is the wavenumber in vacuum, $\omega$ is the light frequency, $c$ is the speed of light,  $\omega_{\scs \mathrm{PBG}}$ is the photonic band gap (PBG) center frequency, and $\lambda_{\scs \mathrm{PBG}}$ is the corresponding wavelength.

Half-wave defect layers, ADL~1 and ADL~2, with the thickness $d_{\mathrm{ADL}} = 2d_e$ are made of the same materials as the anisotropic layer and characterized by the permittivity tensor. The dielectric tensor is determined by the direction of unit vectors 
\begin{equation}
\bm{e_{1,2}} = (\cos{(\phi_{1,2})}, \sin{(\phi_{1,2})}, 0)^{\dagger}.\label{a}
\end{equation}
with reaspect to the coordinate axes.
With given directions of vectors \bm{$e_{1, 2}$}, the permittivity tensor takes the form
\begin{widetext}
\begin{equation}
\hat{\epsilon}_{1,2} = \left\{\begin{array} {cc}
\varepsilon_e \cos^2 (\phi_{1,2}) + \varepsilon_o \sin^2 (\phi_{1,2}) & \sin{(2\phi_{1,2})} \;  (\varepsilon_e-\varepsilon_o)/2 \\
\sin{(2\phi_{1,2})}  \; (\varepsilon_e-\varepsilon_o)/2 & \varepsilon_e \sin^2 (\phi_{1,2}) + \varepsilon_o \cos^2 (\phi_{1,2})
\end{array} \right\}.
\label{eq:epsilon}
\end{equation}
\end{widetext}
The defect layers, ADL~1 and ADL~2, are separated by the PhC containing $2N + 1$ layers, where $N$ is the number of full periods between defects.

All the PhC layers for the $y$-waves have the refractive index $n_o$, while the refractive index for the $x$-waves alternates along the $z$-axis. Therefore, a PBG arises for the $x$-waves only, while the $y$-waves form a continuum of propagating waves. The anisotropic PhC is transparent to $y$-waves and nontransparent for $x$-waves at the normal incidence. The above is illustrated by the spectra in Fig.~\ref{fig1}~(b) calculated by the Berreman transfer matrix method \cite{Berreman1972} with finite number $m$ of periods to the left and to the right of the ADLs. At the angles of rotation of the ADL optical axes $\phi_{1, 2} = 0$, both polarizations are fully decoupled \cite{Timofeev2018_BIC}. Rotating the optical axes of the defect layers, one can mix the two polarizations and localize the $x$-wave. The localization manifests itself in the form of resonant features in the $y$-polarized spectrum in Fig.~\ref{fig1}~(b). One of the resonant features is extremely narrow, which is indicative of a high Q-resonance ($Q \approx 10^6$) to be confirmed by a large amplitude of the localized wave shown in Fig.~\ref{fig1}~(c). It can be seen that the $x$-component is localized within the ADLs similar to an ordinary defect mode. At the same time the $y$-component is localized between ADLs, which is distinctive for the FP-BIC~\cite{HsuChiaWei2016}. 

\section{TCMT Equations for Individual Resonator}

Each ADL can be considered as a resonator and PhCs on both sides as waveguides. The resonators are coupled through the central PhC waveguide. To understand the optical properties of the system of two coupled resonators, we built a fully analytical model based on the temporal coupled-mode theory (TCMT)~\cite{FanShanhui2003, Haus1984bk, Joannopoulos2008bk}.

Let us consider the two-channel scattering. For the incident light polarized along the $y$-axis, the $S$-matrix is implicitly defined by the equation
\begin{equation}\label{direct}
\left(
\begin{array}{c}
s_{1}^{\smallm} \\
s_{2}^{\smallm} \\
\end{array}
\right)
=\widehat{S}_0
\left(
\begin{array}{c}
s_{1}^{\smallp} \\
s_{2}^{\smallp} \\
\end{array}
\right),
\end{equation}
where $s_{m}^{\smallpm}$ are the amplitudes of plane waves in the far-field with a subscript $m=1, 2$ corresponding to the left and right half-spaces and superscripts $^{\smallp}$ and $^{\smallm}$ standing for the incident and outgoing waves, respectively. We assume that the system is illuminated by a monochromatic wave of frequency $\omega$. Below, we introduce the vectors of incident and outgoing amplitudes $|s^{\smallpm}(t)\rangle$, which oscillate in time with the harmonic factor $e^{-i\omega t}$. According to \cite{FanShanhui2003} the TCMT equations take the form
\begin{align}\label{CMT_linear}
& \frac{d a(t)}{d t}=-(i\omega_0+\gamma)a(t)+\langle d^{*}|s^{\smallp}(t)\rangle, \nonumber \\
&|s^{\smallm}(t)\rangle=\widehat{C}_0|s^{\smallp}(t)\rangle+
a(t)|d\rangle, 
\end{align}
where $\widehat{C}_0$ is the matrix of the direct (non-resonant) process, $\omega_0$ is the resonance frequency,
$\gamma$ is the radiation decay rate, $a$ is the amplitude of the resonance eigenmode, and
$|d\rangle$ is the $2\times1$ vector of the coupling constants, which satisfies the conditions
\begin{equation}\label{CMT1}
\langle d|d\rangle=2\gamma,
\end{equation}
\begin{equation}
\widehat{C}_0|d^*\rangle=-|d\rangle.
\end{equation}
The solution for the $S$ matrix is
\begin{equation}\label{S}
\widehat{S}_0=\widehat{C}_0+\frac{|d\rangle\langle d^{*}|}{i(\omega_0-\omega)+\gamma},
\end{equation}
where the direct process matrix is given by 
\begin{equation}\label{C0}
\widehat{C}_0=
e^{i\psi}\left(
\begin{array}{cc}
0 & i\\
i & 0
\end{array}
\right),
\end{equation}
and the coupling vector is 
\begin{equation}\label{d}
    |d\rangle=
    \left(
\begin{array}{c}
d\\
-d
\end{array}
\right), \ d=e^{i\psi/2}\sqrt{\frac{\gamma}{2}}(1+i).
\end{equation}

\section{Two Coupled Closed Resonators}
Let us consider the eigenvalue problem for two coupled BIC at $\phi_{1,2}=0$
\begin{equation}
    \widehat{\mathcal{L}}\mathbb{E}_0=-i\omega_0\widehat{\mathcal{E}}_0\mathbb{E}_0,
\end{equation}
where
\begin{equation}
    \widehat{\mathcal{L}}=
    \left(
    \begin{array}{cc}
        0 & \nabla\times \\
        -\nabla\times & 0
    \end{array}
    \right)
\end{equation}
is the Maxwell operator, $\omega_0$ is the eigenfrequency,
\begin{equation}
    {\mathbb{E}}_0=
    \left(
    \begin{array}{c}
        \bm{E} \\
        \bm{H}
    \end{array}
    \right)
\end{equation}
is the eigenvector, and
\begin{equation}
    \widehat{\mathcal{E}_0}=
    \left(
    \begin{array}{cc}
        \hat{\epsilon} & 0 \\
        0 &  \widehat{I}
    \end{array}
    \right).
\end{equation}

The eigenvector ${\mathbb{E}}_0$ corresponds to the symmetry-protected BIC \cite{Timofeev2018_BIC}. 
Now, we consider two identical resonators separated by a PhC waveguide, each supporting a BIC as shown in Fig.~\ref{fig1}~(d). Since the symmetry is not broken, the resonators are only coupled via the evanescent tails of the BIC eigenmodes due to the tunneling across the PBG. Each eigenmode is a solution of the source-free Maxwell equation with the permittivity tensor corresponding to each single resonator
\begin{equation}
   \widehat{\mathcal{L}}\mathbb{E}_{1,2}=-i\omega_0\widehat{\mathcal{E}}_{1,2}\mathbb{E}_{1,2}. 
\end{equation}
The eigenmodes $\mathbb{E}_{1, 2}$ are localized at each individual resonator.
The temporal Maxwell equations
\begin{equation}\label{problem}
     \widehat{\mathcal{L}}\mathbb{E}=
     \widehat{\mathcal{E}}\frac{d\mathbb{E}}{dt}
\end{equation}
should be solved with the new permittivity tensor, see Fig.~1 in Supplementary Materials,
\begin{equation}
    \widehat{\mathcal{E}}=
    \left\{ 
    \begin{array}{c} 
    \widehat{\mathcal{E}}_1 \ \ \mathrm{if}  \ \ z<0 \\
    \widehat{\mathcal{E}}_2 \ \ \mathrm{if} \ \ z>0
    \end{array}
    \right.
\end{equation}
Let us find the solution in the form
\begin{equation}\label{solution}
    \mathbb{E}(t)=a_1(t)\mathbb{E}_{1}+a_2(t)\mathbb{E}_{2}.
\end{equation}
Substituting \eqref{solution} into \eqref{problem} and using the normalization condition
\begin{equation}
    \int dz\mathbb{E}^{\dagger}_{1,2}\widehat{\mathcal{E}}_{1,2}\mathbb{E}_{1, 2} = 1,
\end{equation}
we can find
\begin{equation}
    -i\omega_0
    \left(
    \begin{array}{cc}
       1  & I_1 \\
        I_1 & 1
    \end{array}
    \right)
    \left(
    \begin{array}{c}
       a_1 \\
        a_2
    \end{array}
    \right)=
    \left(
    \begin{array}{cc}
       1  & I_2 \\
        I_2 & 1
    \end{array}
    \right)\frac{d}{dt}
    \left(
    \begin{array}{c}
       a_1 \\
        a_2
    \end{array}
    \right),
\end{equation}
where
\begin{equation}
    I_{1}=\int dz \mathbb{E}^{\dagger}_{1}\widehat{\mathcal{E}}_2\mathbb{E}_{2}=
    \int dz \mathbb{E}^{\dagger}_{2}\widehat{\mathcal{E}}_1\mathbb{E}_{1},
    \label{I1}
\end{equation}
and
\begin{equation}
    I_{2}= \int dz \mathbb{E}^{\dagger}_{1}\widehat{\mathcal{E}}\mathbb{E}_{2} = \int dz \mathbb{E}^{\dagger}_{2}\widehat{\mathcal{E}}\mathbb{E}_{1}.
        \label{I2}
\end{equation}
Taking into account 
\begin{equation}
    I_{1, 2}\ll1,
\end{equation}
we arrive at
\begin{equation}
    \left(
    \begin{array}{cc}
         \omega_0  & v \\
        v &   \omega_0
    \end{array}
    \right)
    \left(
    \begin{array}{c}
       a_1 \\
        a_2
    \end{array}
    \right)=
    i\frac{d}{dt}
    \left(
    \begin{array}{c}
       a_1 \\
        a_2
    \end{array}
    \right),
\end{equation}
where the tunneling coupling constant is
\begin{equation}\label{v}
    v=\omega_0(I_1-I_2).
\end{equation}

\section{Two Resonators Coupled with Two Waveguides}

\begin{figure*}[t]
\center{\includegraphics{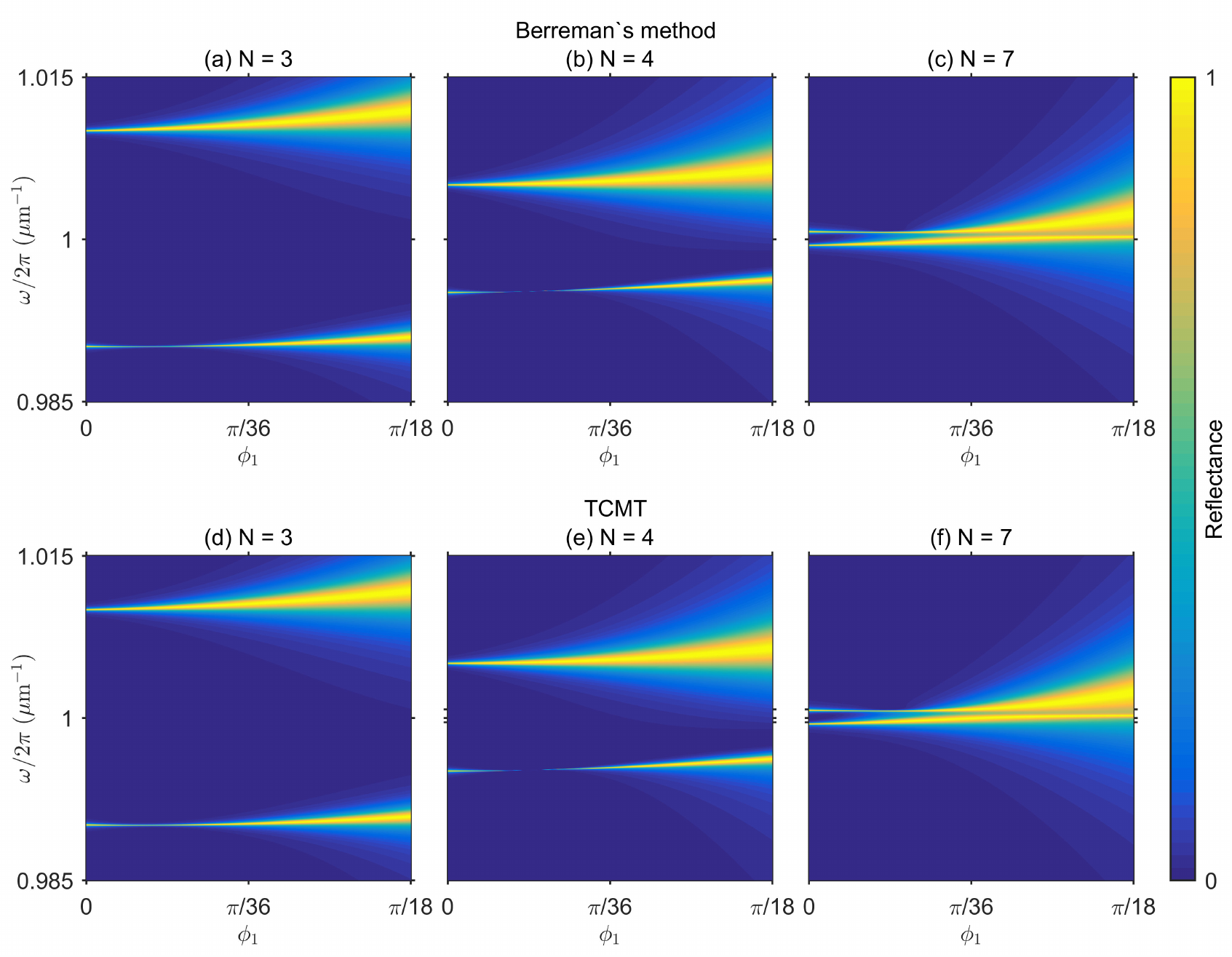}}
\caption{Reflectance spectra obtained using (a--c) the Berreman transfer matrix method and (d--f) the TCMT model at $N = 3$ (a, d); $N = 4$ (b, e), and $N = 5$ (c, f). The other parameters are the same as in the caption to Fig.~\ref{fig1}.}
\label{fig2}
\end{figure*}

\begin{figure*}[t]
\center{\includegraphics{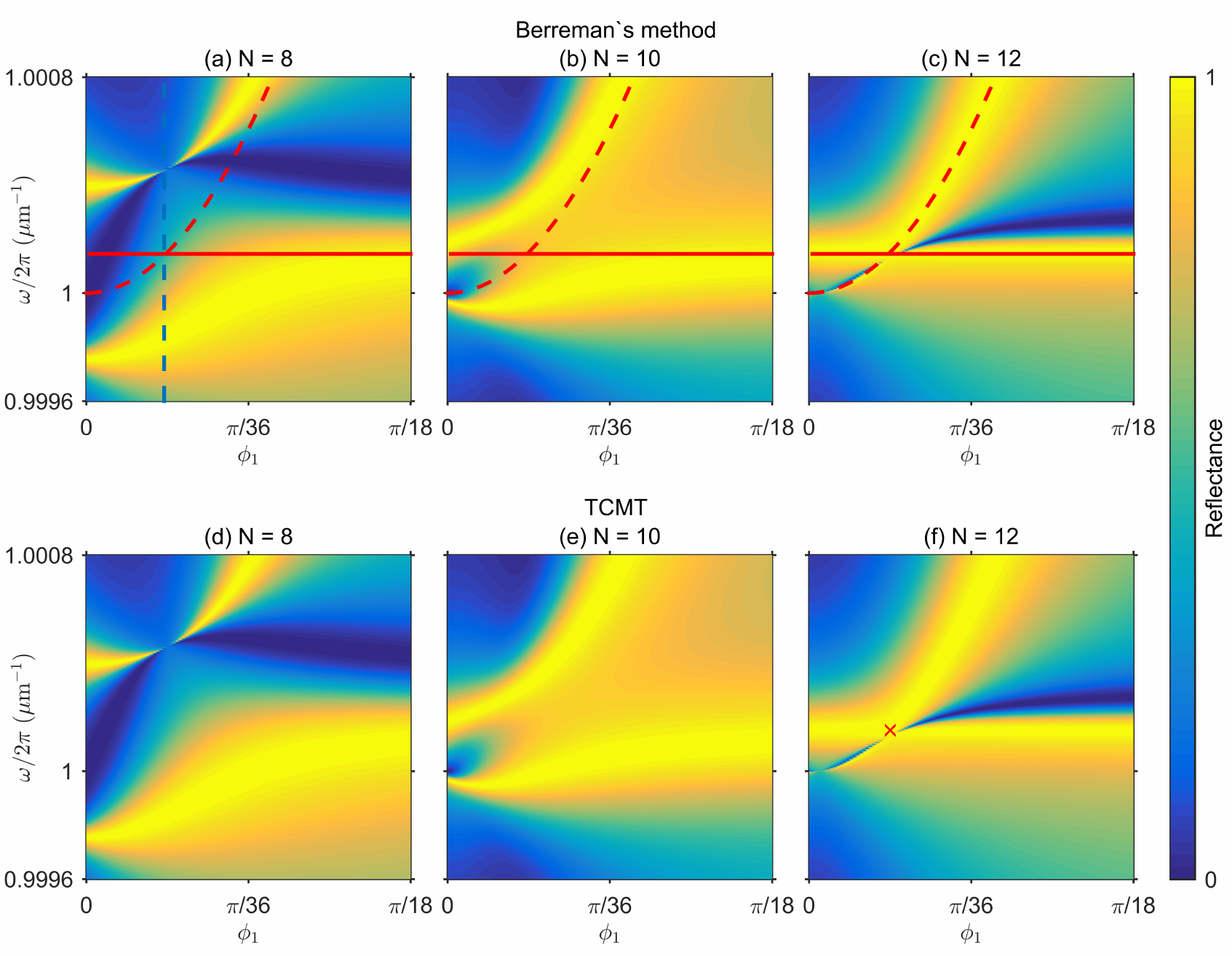}}
\caption{Reflectance spectra obtained using (a--c) the Berreman transfer matrix method and (d--f) the TCMT model at $N = 8$  (a, d), $N = 10$ (b, e) , and $N = 12$ (c, f). The other parameters are the same as in the caption to Fig.~\ref{fig1}. The blue dashed line in (a) at an angle of $\phi_1 = 2.4\pi/180$ corresponds to the spectrum in Fig.~\ref{fig1}~(b). The red lines in (a--c) correspond to eigenfrequencies $\omega_{1}$ (dashed line) and $\omega_{2}$ (solid line) obtained from Eq.~\eqref{omega}. The red cross corresponds FP-BIC friquency position obtained from Eq.~\eqref{v_BIC_1}, and Eq.~\eqref{v_BIC_2}.}
\label{fig3}
\end{figure*}

\begin{figure*}[t]
\center{\includegraphics{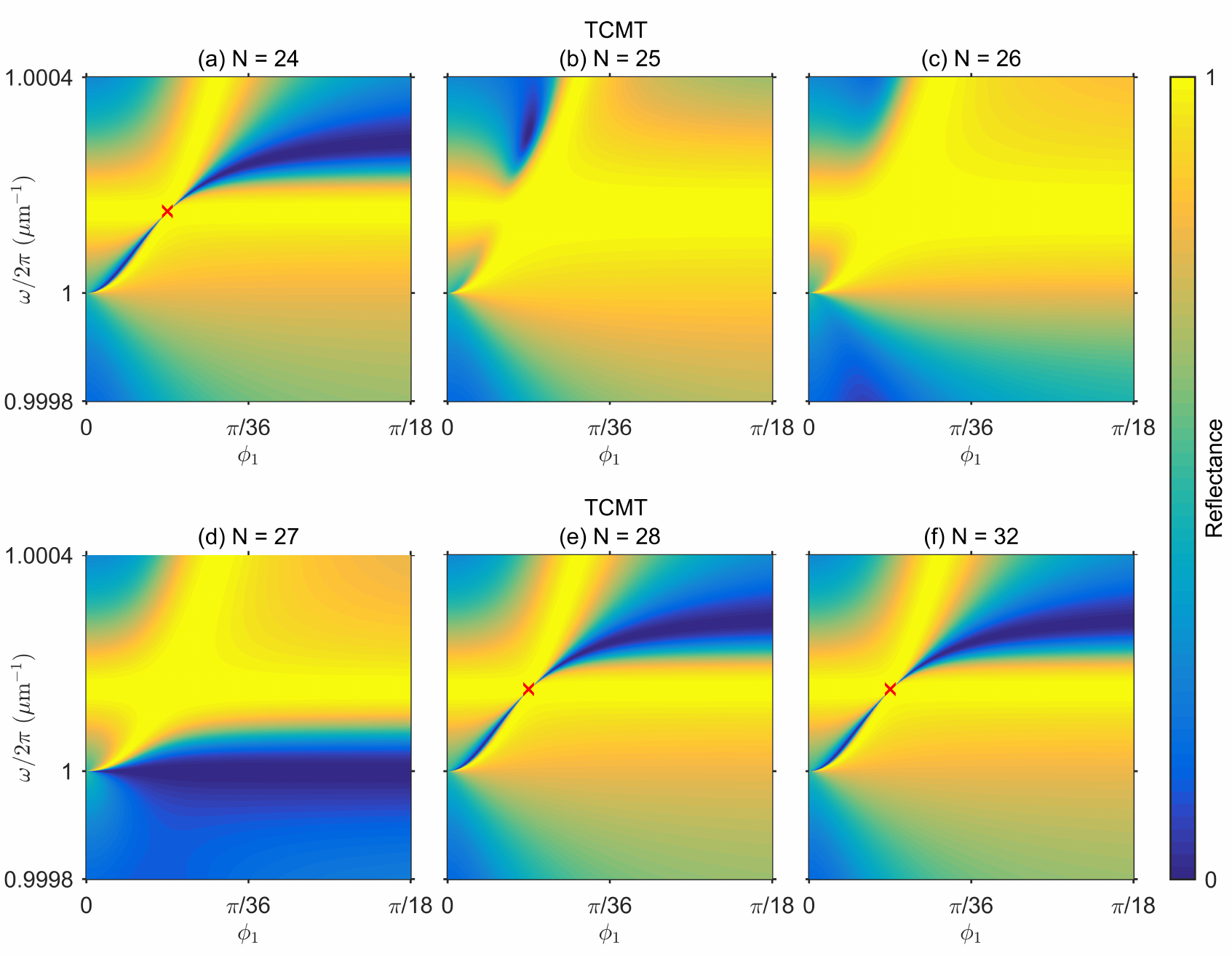}}
\caption{Reflectance spectra obtained within the TCMT model at $N$ = 24  (a), 25 (b), 26 (c), 27 (d), 28 (e), and 32 (f). The rest parameters are the same as in the caption to Fig.~\ref{fig1}. Red crosses correspond FP-BIC friquency positions obtained from Eq.~\eqref{v_BIC_1}, and Eq.~\eqref{v_BIC_1}.}
\label{fig4}
\end{figure*}

Let us now consder the case $\phi_{1,2}\neq0$.The BICs in the ADLs now became
quasi-BIC, i.e. the resonantors are now coupled with $y$-waves.
The TCMT equation for amplitudes of the resonant modes is
\begin{widetext}
\begin{equation}\label{two_modes}
   \left(
    \begin{array}{cc}
          i(\omega_1-\omega)+\gamma_1 & iv \\
         iv & i(\omega_2-\omega)+\gamma_2 
    \end{array}
    \right)
\left(
\begin{array}{c}
    a_1  \\
      a_2
\end{array}
\right)=
 \left(
\begin{array}{c}
    d_1(s_1^{\smallp}-s_{\scs{\leftarrow}}e^{ikL}) \\
    d_2(s_2^{\smallp}-s_{\scs{\rightarrow}}e^{ikL})
\end{array}
\right),   
\end{equation}
\end{widetext}
where $L$ is the distance between the resonators as shown in Fig.~\ref{fig1}~(d).
For the reflected waves, we have
\begin{align}\label{reflected}
& s_1^{\smallm}=d_1a_1+ie^{i(\psi+kL)}s_{\scs{\leftarrow}}, \nonumber \\
& s_2^{\smallm}=d_2a_2+ie^{i(\psi+kL)}s_{\scs{\rightarrow}}.
\end{align}
At the same time, for the waves between the resonators, we can write
\begin{align}\label{between}
& s_{\scs{\rightarrow}}=-d_1a_1+ie^{i\psi}s^{\smallp}_1, \nonumber \\
& s_{\scs{\leftarrow}}=-d_2a_2+ie^{i\psi}s^{\smallp}_2.
\end{align}
Combining Eq.~\eqref{reflected} and Eq.~\eqref{between}, we find
\begin{align}\label{reflected2}
& s_1^{\smallm}=d_1a_1-id_2e^{i(\psi+kL)}a_2 -e^{i(2\psi+kL)}s_2^{\smallp},
\nonumber \\
& s_2^{\smallm}=d_2a_2-id_1e^{i(\psi+kL)}a_1 -e^{i(2\psi+kL)}s_1^{\smallp}
.
\end{align}
Substituting Eq. \eqref{between} into Eq.~\eqref{two_modes}, we obtain
\begin{widetext}
\begin{equation}\label{two_mode}
   \left(
    \begin{array}{cc}
          i(\omega_1-\omega)+\gamma_1 & iv-d_1d_2e^{ikL} \\
         iv-d_1d_2e^{ikL} & i(\omega_2-\omega)+\gamma_2 
    \end{array}
    \right)
\left(
\begin{array}{c}
    a_1  \\
      a_2
\end{array}
\right)=
 \left(
\begin{array}{c}
    d_1(s_1^{\smallp}-is_{2}^{\smallp}e^{i(kL+\psi)}) \\
    d_2(s_2^{\smallp}-is_{1}^{\smallp}e^{i(kL+\psi)})
\end{array}
\right).   
\end{equation}
The above set of equations can be solved for $a_1$ and $a_2$.
The reflection amplitude is then found from Eq. \eqref{reflected2}:

\begin{equation}\label{r}
     \rho = |\widehat{A}|^{-1}(d_1^2(i(\omega_{2} - \omega) + \gamma_2) +2id_1d_2e^{i(\psi+kL)}(iv - d_1d_2e^{ikL}) + d_2^2e^{2i(\psi+kL)}(i(\omega_{1} - \omega) - \gamma_1),
     \end{equation}
\end{widetext}
where
\begin{equation}\label{A}
\widehat{A} = \left(
    \begin{array}{cc}
          i(\omega_1-\omega)+\gamma_1 & iv-d_1d_2e^{ikL} \\
         iv-d_1d_2e^{ikL} & i(\omega_2-\omega)+\gamma_2 
    \end{array}
    \right).
\end{equation}

In the case of a single defect layer, direct process matrix \eqref{C0} has the form
\begin{equation}
\widehat{C}_0 = e^{ik_od_{ADL}}\left(
\begin{array}{cc}
0 & 1\\
1 & 0
\end{array}
\right),
\end{equation}
i.e., the $y$-wave, propagating through the defect layer, accumulates the phase $k_od_{\mathrm{ADL}}$, then
\begin{equation}
\psi = k_od_{\mathrm{ADL}} - \pi/2.
\end{equation}
Then, the coupling constant \eqref{d} takes the form
\begin{equation}
d_{1,2} = e^{ik_od_{\mathrm{ADL}}/2}\sqrt{\gamma_{1,2}}.
\label{d_12}
\end{equation}
The expressions for $\omega_{1, 2}$ and $\gamma_{1, 2}$ were obtained in \cite{Pankin2020Fano}:
\begin{equation}
\begin{aligned}
& \omega_{1,2} = \omega_{\scs \mathrm{PBG}} + \frac{\omega_{\scs \mathrm{PBG}}}{\pi} q (1 - q) \sin{(\pi q)} \cdot \phi_{1,2}^{2} + {\cal O}(\phi_{1,2}^4),\\
& \gamma_{1,2} = \frac{2 \omega_{\scs \mathrm{PBG}}}{\pi} q (1 - q) \cos^2{(\pi q/2)} \cdot \phi_{1,2}^{2}+ {\cal O}(\phi_{1,2}^4),
\label{omega}
\end{aligned}
\end{equation}
where $q = n_{o}/n_{e}$.
The equations for ${\mathbb{E}}_{1, 2}$ were derived in \cite{Timofeev2018_BIC}, see Eqs.~(8-12) in the latter reference. In Fig.~1 in Supplementary Materials we plotted the field ${\mathbb{E}}_{1, 2}$ distributions.
The phase $kL$ \eqref{two_modes} accumulated by the $y$-wave in propagation between ADL~1 and ADL~2, see Fig.~\ref{fig1}~(d) is 
\begin{equation}\label{L}
kL = k_oN(d_o + d_e) + k_od_o.
\end{equation}

\section{Results and Discussion}

Figures~\ref{fig2}-\ref{fig4} show the reflectance spectra calculated by the Berreman transfer matrix method and the TCMT. It can be seen that two resonant lines approach each other with an increase in number of periods $N$ in the PhC between the ADLs. At $N = 4,8,12,16,20,24,28,32$ an so on, the width of one of the of resonant lines collapses, if $\phi_1 = \phi_2 = \pi/72$.

The collapses of the resonant lines result from the coupling between the resonant modes localized in both ADLs, which is evidenced by the avoided crossing. 
In Figs.~\ref{fig3}~(a-c), the red dashed line shows the resonant frequency $\omega_1$ as a function of the rotation angle for the structure containing only ADL~1. The red solid line shows the resonant frequency $\omega_2$ for the structure containing only ADL~2, the rotation angle of which is fixed. Both lines are obtained using Eqs.~\eqref{omega}. It can be seen that, in the system with two ADLs, the resonant lines pass below and above the resonance frequencies $\omega_{1, 2}$. The coupling between the resonant modes is due to the off-diagonal elements of the matrix $\widehat{A}$, Eq.~\eqref{A}, see \cite{limonov2017fano} for more detail.
The tunneling coupling constant $v$ \eqref{v} tends to zero with the increase of the number of periods between the ADLs
\begin{equation}\label{limes}
\lim_{N \to \infty} \limits v = 0,
\end{equation}
since the field distributions ${\mathbb{E}}_{1, 2}$ are evanescent functions decaying exponentially outside the ADL \cite{das2020resonant}, i.e. $I_{1, 2} = 0$ in Eqs.~(\ref{I1}-\ref{I2}), see Fig.~1 in Suplementary Materials. This explains the repulsion of the resonant lines with decreasing $N$.

It can be seen in Fig.~\ref{fig4} that at large $N$ the spectra replicate, see Fig.~\ref{fig4}~(a, e, f), with a period of $\Delta N = 4$. This can be explained by the fact that in Eq~\eqref{r} 
we can ignore the terms that include $v$ at large $N$. The resulting equation does not change with an increase in $L$ by an integer number $\ell$ of half-waves 
\begin{equation}
\rho(kL) = \rho(kL + \ell\pi).
\end{equation}
According to Eq.~\eqref{L} it can be shown that with our calculation parameters $k_od_o = \pi/2$ and $k_od_e = \pi/4$ for $\omega = \omega_{\mathrm{PBG}}$, the smallest integer is $\ell = 3$ at $\Delta N =  4$.
The spectra obtained with the TCMT, Fig.~\ref{fig2}, and Fig.~\ref{fig3}~(d-e) are consistent with the spectra obtained by the Berreman method, Fig.~\ref{fig2}, and Fig.~\ref{fig3}~(a-c), also see Fig.~2 in Suplementary Materials. The difference between the two methods is observed when the approximations used to build the TCMT, $\phi_{1,2} \ll 1$ and $v \ll 1$, break down, see Fig.~3 in Suplementary Materials.

The parameters at which the resonant line collapses can be found by solving the eigenvalue problem, which is formulated as
\begin{equation}\label{BIC}
\widehat{A} |a\rangle = 0.      
\end{equation}
The BIC can be found as a solution of Eq.~\eqref{BIC} with a real eigenfrequency $\omega = \omega_{\scs \mathrm{BIC}}$. 
However, there is a more convenient way to obtain the FP-BIC condition for $N \gg 1$.
The two ADLs can act as a pair of perfect mirrors that trap waves between them. The FP-BICs are formed when the resonance frequency or the spacing, $L$, between the two ADLs is tuned to make the round-trip phase shifts add up to an integer multiple of $2\pi$~\cite{HsuChiaWei2016}.
Then, the equation for the FP-BICs have the following form:
\begin{equation}
\psi_{\mathrm{res}} + k_oL = \pi \ell,
\label{FP_equation}
\end{equation}
where $\psi_{\mathrm{res}} = arg(S_{11})$ found from Eq.~\eqref{S} is the phase of the resonant reflection from the ADL, and $\ell$ is an integer number.
From the Eq.~\eqref{d_12}
 $\psi_{\mathrm{res}}$ has the following form:
\begin{equation}
\psi_{\mathrm{res}} = 2 \psi_d = \psi + \pi/2 = k_od_{\scs \mathrm{ADL}},
\end{equation}
where $\psi_d = arg(d)$ is the phase of the coupling constant Eq.~\eqref{d}.
Taking into account that $k_o = \omega_{\scs \mathrm{BIC}}n_o/c$ and the expression for $L$ \eqref{L} we can obtain the equation for the FP-BICs frequencies
\begin{equation}\label{v_BIC_1}
\omega_{\scs \mathrm{BIC}} = \frac{\pi \ell c}{n_o(d_{\scs \mathrm{ADL}} + N(d_o + d_e) + d_o)}.
\end{equation}
The required value $\ell$ is defined using the equation \eqref{omega} as follows
\begin{equation}\label{v_BIC_2}
\omega_{\scs \mathrm{BIC}} = \omega_{1,2}(\phi_{1,2}).
\end{equation}
The solution of Eq.~\eqref{v_BIC_1}, and Eq.~\eqref{v_BIC_2} is shown im Fig.~\ref{fig3}, and Fig.~\ref{fig4} by red crosses. It can be seen that for $N = 12$, Fig.~\ref{fig3}~(f), the FP-BIC frequency found from the above equations matches the numerical data to a good accuracy, while for $N = 24, 28, 30$ it corresponds the exact position of the resonant line collapse. The deviation at $N=12$ is because Eq.~\eqref{FP_equation} neglects the tunneling coupling constant $v$, Eq.~\eqref{limes}.   

\section{Conclusions}

In this work, the Fabry--Perot BICs are found in an anisotropic photonic crystal containing two anisotropic defect layers. Each defect layer can separately support a symmetry-protected BIC, thereby acting as an ideal mirror in the Fabry--Perot resonator. A fully analytic model is proposed to solve the scattering problem within the framework of the temporal coupled-mode theory. The spectra found using the analytic model are consistent with the numerical spectra obtained using the Berreman transfer matrix method. The analytic model explains the spectral features, in particular, the avoided crossing of the resonant lines, collapses of the resonant lines in the Fabry--Perot BIC points, and the periodicity of the spectra in the case of defect layers at a large distance from one another. The proposed model can be used to design microcavities with controllable Q-factor \cite{pankin2020one, wu2021quasi}.         

\section*{Acknowledgments}
We acknowledge discussions with Almas F. Sadreev.
This study was supported by the Council on Grants of the President of the Russian Federation (MK-4012.2021.1.2).


%

\end{document}